# Effect of adding nanometre-sized heterogeneities on the structural dynamics and the excess wing of a molecular glass former

S. Gupta,[1,2,3*] J. K. H. Fischer,[4] P. Lunkenheimer,[4] A. Loidl,[4] E. Novak,[5] N. Jalarvo,[1,6] and M. Ohl[1,2]

[1]*Juelich Centre for Neutron science (JCNS) outstation at SNS, POB 2008, 1 Bethel Valley Road, TN 37831, Oak Ridge, USA*

[2]*Biology and Soft Matter Division, Neutron Sciences Directorate, Oak Ridge National Laboratory (ORNL), POB 2008, 1 Bethel Valley Road, TN 37831, Oak Ridge, USA*

[3]*Department of Chemistry and Macromolecular Studies Group, Louisiana State University, Baton Rouge, LA 70803, USA*

[4]*Experimental Physics V, Center for Electronic Correlations and Magnetism, University of Augsburg, 86135 Augsburg, Germany*

[5]*Department of Materials Science and Engineering, University of Tennessee, Knoxville, TN 37996, USA*

[6]*Chemical and Engineering Materials Division, Oak Ridge National Laboratory (ORNL), PO BOX 2008 MS6473, Oak Ridge, TN 37831, USA*

*Correspondence to S.G. (email: sudiptag@lsu.edu)

We present the relaxation dynamics of glass-forming glycerol mixed with 1.1 nm sized polyhedral oligomeric silsesquioxane (POSS) molecules using dielectric spectroscopy (DS) and two different neutron scattering (NS) techniques. Both, the reorientational dynamics as measured by DS and the density fluctuations detected by NS reveal a broadening of the $\alpha$ relaxation when POSS molecules are added. Moreover, we find a significant slowing down of the $\alpha$-relaxation time. These effects are in accord with the heterogeneity scenario considered for the dynamics of glasses and supercooled liquids. The addition of POSS also affects the excess wing in glycerol arising from a secondary relaxation process, which seems to exhibit a dramatic increase in relative strength compared to the $\alpha$ relaxation.

**Introduction**

The glass and supercooled-liquid states of matter are found in a large variety of materials and there are many applications of glass such as windows and containers, optical fibres, food processing or even preservation of insect life under extreme conditions. An ever-increasing industrial demand and critical role in economic sustainment require the development of a basic understanding of such materials. Enormous efforts have been made during the last decades to arrive at a better understanding of the glassy state of matter. (In this work, the term "glassy" refers to both the supercooled-liquid and the glass state, above and below the glass temperature, respectively.) Nevertheless, the glass transition is still one of the most interesting unsolved problem in solid state physics. Above the glass-transition temperature $T_g$, the glassy dynamics is dominated by the structural $\alpha$ relaxation but there are also various faster relaxation processes, which seem to be universal properties of the glassy state of matter[1–5]. The microscopic origin of these faster relaxations is still a highly debated topic in modern science. In broadband dielectric spectroscopy (DS) the $\alpha$ relaxation leads to a dominant peak in the dielectric loss ($\varepsilon''$) spectra[3,4]. In various glass formers, it is followed by a broad second peak or shoulder arising from a secondary relaxation, termed slow $\beta$ or Johari-Goldstein (JG) relaxation[5], which becomes best visible when approaching $T_g$ (refs. 5–7). One of the most mysterious phenomena among the processes beyond the $\alpha$ relaxation is the excess wing (EW)[3,4,8–12] whose physical origin still is very controversial. It is typically observed in DS as a second, more shallow power law at the high-frequency flank of the $\alpha$ peak of various glass formers. The EW is often assumed to arise from a JG relaxation peak, partially submerged under the dominating $\alpha$-peak[13–15] but other interpretations were also proposed[6,16]. The JG or EW processes occur even for molecules without internal degrees of freedom, pointing towards an intermolecular origin[5]. One possible explanation of the JG relaxation is the existence of "islands of mobility", i.e., regions where the molecules have greater mobility[5,17]. An alternative explanation assumes that *all* molecules contribute to the JG relaxation via transitions between local minima within the energy landscape experienced by the molecules[18,19]. Overall, a proper understanding of these processes is still missing and often regarded as prerequisite to achieve a better understanding of the glass transition in general.

It should be noted that the JG relaxation and the EW are distinctively different from the well-known fast $\beta$ process as predicted by mode-coupling theory (MCT), which has proven to be successful in explaining some of the most interesting phenomena of glassy dynamics[20]. According to MCT, the fast $\beta$ process is related to the molecular dynamics in the transient "cage" formed by the neighbouring molecules whereas the low-frequency $\alpha$ process is the relaxation mode by which the cages form and decay. However, an explanation for the EW or the JG (= slow $\beta$) process is missing within the original MCT framework (however, see ref. 21 for a possible explanation of these processes within more recent MCT concepts).

The random nature of molecular diffusion in liquids causes local fluctuations of intermolecular distances. It gave birth to the concept of *dynamic heterogeneities*[22–25] or spatially heterogeneous dynamics[26]. It leads to a distribution of relaxation times, i.e., each molecule relaxes with exponential time dependence as in the Debye theory, but the relaxation times are different for different molecules. This explains one of the hallmark features of glassy dynamics, the non-exponentiality, which shows up, e.g., by a broadening of the $\alpha$ peaks in dielectric-loss spectra. It has been recently observed for soft colloidal glasses that on suppressing the dynamic heterogeneities as a result to its inherent intermolecular softness one can attain a hyperuniform state on approaching the glass transition[27], in contrast to hard colloids where dynamic heterogeneity dominates[24].

In the present work, following the previous work by Johari and Khouri[28,29] we investigate the influence of structural heterogeneity on molecular dynamics by adding POSS molecules with a size comparable to that of typical dynamic heterogeneous domains of 1-3 nanometers[30,31] to the classical molecular glass former glycerol. We investigate the influence of these heterogeneities on the $\alpha$ relaxation and, in addition, also explore possible effects on the EW. We use static and dynamic NS as well as DS to determine the static single-particle structure and heterogeneity-induced variations of the relaxational dynamics. The added hybrid-plastic POSS molecules consist of a silicon-oxygen cage ($Si_8O_{12}$) in which the Si atoms occupy the corners of a cube and are bonded to oxygen atoms which are located on the edges (cf. Fig. 1). A wide variety of ligands can be bound to the fourth position of the Si atom in the cage. The intermolecular reactivity of these ligands allows the molecules to be dispersed in various complex fluids[30]. Indeed, the influence of added POSS molecules on glassy dynamics has been investigated in several previous works[28,29,33,34].

## Results

**Small angle neutron scattering (SANS).** Figure 1 shows the SANS intensity of a deuterated (d) p-xylene solution of Octa(3-hydroxy-3-methylbutyldimethulsiloxy) POSS (POSS-OH) with a volume fraction of $\phi_{POSS} = 1\%$ as a function of the scattering vector, $Q$. The solid lines are fits using a hard sphere (HS) form factor given by[35]:

$$\frac{d\Sigma}{d\Omega}(Q) = \frac{\phi_{POSS}}{V} \Delta\rho^2 \left[ 3V \frac{\sin(Qr) - (QR)\cos(Qr)}{(Qr)^3} \right]^2 \qquad (1)$$

Here, $V$ is the volume of the POSS molecule, $r$ is the hard sphere radius, and $\Delta\rho = \rho_{POSS} - \rho_{solvent}$ is the contrast difference between POSS and the solvent. It results in a radius of $r = 11\pm0.1$Å, which is about ~7 times the hydrodynamic radius ($R_H$) of a glycerol molecule (1.6 Å)[36]. The inset in Fig. 1 illustrates the molecular structure of completely condensed POSS-OH[37]. Its molecular formula is $C_{56}H_{136}O_{28}Si_{16}$, with the ligand group $R = SiO_2C_7H_{17}$. The SANS data was modelled using 0% polydispersity.

**Neutron Backscattering Spectroscopy (BS).** In Fig. 2 we report the incoherent quasi-elastic NS (QENS) data from BASIS BS[38] for an average momentum transfer, at (a) $<Q> = 1.3\pm0.2$ and (b) $1.5\pm0.2$ Å$^{-1}$, in a semi-logarithmic representation. It is plotted over a temperature range from 273 K to 413 K for 7.7 wt% of POSS NPs in glycerol-$H_8$. At low temperatures ($T \leq 294$ K) the BS data were acquired at the edge of the instrument resolution but with extended measurement time to obtain improved statistics. The QENS intensity is proportional to the incoherent dynamic structure factor, $S_{inc}(Q,E)$, which contains correlations between the same nucleus at different times (incoherent scattering) and between different nuclei at different times (interference effects associated with coherent scattering, both elastic and inelastic). The solid black lines in Fig. 2 (a-b) represent the modelling of the BS data. All obtained BS results were fitted for each $Q$ value (over a $Q$ range: 0.3 - 1.9 Å$^{-1}$) using the following expression[9]:

$$S_{inc}(Q,E) = A\{X(Q)\delta(E) + [1 - X(Q)]S_{qse}(Q,E)\} \otimes R(Q,E) + B(Q,E) \qquad (2)$$

Here $A \sim \exp\left(-\langle\Delta r^2\rangle Q^2\right)$ is the Debye-Waller factor taking into account the decay of the correlations at faster times; it depends exponentially on $\langle\Delta r^2\rangle$, the effective mean-square-displacement (MSD) in a quasi-harmonic approximation[39]. $X(Q)\delta(E)$ represents the fraction of the elastic scattering, $[1 - X(Q)]S_{qse}(Q,E)$ are the quasi-elastic contribution, and $B(Q,E)$ is a linear background term. The term $X(Q)$ also acts as the elastic incoherent structure factor (EISF). $B(Q,E)$ represents any elastic background arising from the sample holders, sample environment, and the spectrometer background. The corresponding resolution function $R(Q,E)$ (dashed line in Fig. 2) was measured for the sample at low temperature (20 K); the symbol $\otimes$ denotes a convolution function. The asymmetry in $S_{inc}(Q,E)$ in Fig. 2 (a-b) is an inherent property of the time-of-flight operation of the BASIS instrument. It is related to the arrival time of fast and slow neutrons in the BS spectrometer[38].

$S_{qse}(Q, E)$ is modelled using the Fourier transformed Kohlrausch-Williams-Watts (KWW) function[40,41], given in time domain by:

$$S_{qse}(Q, E) = \int_0^\infty dt\, e^{i\omega t} \exp\left[-\left(\frac{t}{\tau_{KWW}}\right)^{\beta_{KWW}}\right] \tag{3}$$

The stretching exponent $\beta_{KWW}$ ($0 < \beta_{KWW} \leq 1$) and the relaxation time $\tau_{KWW}$ are used to determine the average relaxation time $\langle \tau_{KWW} \rangle = \tau_{KWW}/\beta_{KWW}\Gamma(1/\beta_{KWW})$, where $\Gamma$ is the Gamma function. Here, $\omega$ denotes the angular frequency ($= E/\hbar = 2\pi\nu$). As can be seen from Fig. 2 (a-b) at low temperature ($T \leq 294$ K) the QENS signal is dominated by the resolution function, the effect of which is visible in the ripples in the energy transfer[42]. We analyse $\langle \tau_{KWW} \rangle$ at $\langle Q \rangle = 1.3$ and 1.5 Å$^{-1}$ for the further course of this work, which is in close vicinity to the coherent static structure factor peak of pure glycerol ($Q_0 = 1.44$ Å$^{-1}$)[43]. This ensures that the observed dynamics is at the length scale before and after the structural relaxation ($Q_0$) of pure glycerol. The obtained stretching exponent $\beta_{KWW}$ varies between 0.48 and 0.56 over a temperature range from 413 K to 323 K (Fig. 3).

As revealed by the relaxation-time analysis (see discussion section), below 323 K the BS data are dominated by the secondary relaxation mode generating the EW. Thus at these temperatures they are modelled using a Cole-Cole (CC)[44] function in energy space, replacing $S_{qse}(Q, E)$ in equation 2 (which is proportional to the loss $\chi''$) by[45]:

$$S_{qse}(Q, E) = \frac{\left(\exp\left(\frac{E}{k_B T}\right) - 1\right)^{-1} \cos\left(\frac{\alpha\pi}{2}\right)\left(\frac{E}{E_0}\right)^{1-\alpha}}{1 + 2\sin\left(\frac{\alpha\pi}{2}\right)\left(\frac{E}{E_0}\right)^{1-\alpha} + \left(\frac{E}{E_0}\right)^{2-2\alpha}} \tag{4}$$

Here $E_0$ is the energy corresponding to the inverse relaxation time, with $\alpha$ ($0 \leq \alpha < 1$) the broadening parameter and $k_B$ the Boltzmann constant. The pre-factor $\left(exp\left(\frac{E}{k_B T}\right) - 1\right)^{-1}$, represents the Bose factor. The CC function is known to provide a good description of secondary relaxations[7,15]. The red solid lines in Fig. 2 represent the CC fits for 273 K and 294 K. We obtain $\alpha = 0.75 \pm 0.01$, which corresponds to a strong broadening of the relaxation peak leading to the EW in the DS data [4,13,14] (see also discussion of the DS data below).

Fig. 2(c) shows the EISF from equation (2) that represents the relative strength of the quasi-elastic translational and reorientational dynamics with respect to the elastic part[46]. The solid lines represent the fits for diffusion in a sphere of radius $R_{trans}$ modelled as, $(3j_1(QR_{trans})/(QR_{trans}))^2$, (translational)[39,46]. The dashed lines represent the convolution of the localized translational centre-of-mass motion and segmental reorientational motion, $[c_1 + (1-c_1)(3j_1(QR_{trans})/(QR_{trans}))^2]\left[c_2 + (1-c_2)\left(j_0{}^2(QR_{rot})\right)\right]$[46]. Here $j_0$ and $j_1$ are the spherical Bessel functions. The parameter $c_1 = 0.94$ was obtained that represents short range bond ordering or heterogeneity. The term, $j_0{}^2(QR_{rot})$[39], represents the rotational diffusion jump on a sphere of radius $R_{rot}$. The parameter $c_2 = 0.66$ was obtained that represents the fraction of H-atoms not participating in the reorientational motion. The two vertical lines indicate $\langle Q \rangle = 1.3$ and $1.5$ Å$^{-1}$. The variation of the confinement lengths with temperature are illustrated in Fig. 2 (d).

**Neutron Spin Echo (NSE) Spectroscopy.** In Fig. 4 we report the normalized dynamic structure factor ($S(Q,t)/S(Q,0)$) of deuterated glycerol-D$_8$, measured by coherent NSE spectroscopy over a temperature range from 273 to 323 K and at $\langle Q \rangle = 1.3 \pm 0.2$ Å$^{-1}$. The data for 7.7 wt% of POSS in glycerol-D$_8$ (open symbols) are compared to those obtained for pure glycerol-D$_8$ (closed symbols). Because of the spin flip at the NSE instrument[47], we detect the normalized total signal representing the sum of coherent ($S_{coh}$) and incoherent ($S_{inc}$) scattering. However, the coherent signal dominates, i.e.,

$$\frac{S(Q,t)}{S(Q,0)} = \frac{S_{coh}(Q,t) - \frac{1}{3}S_{inc}(Q,t)}{S_{coh}(Q,0) - \frac{1}{3}S_{inc}(Q,0)} \approx \frac{S_{coh}(Q,t)}{S_{coh}(Q,0)} \tag{5}$$

This coherent intermediate scattering function obtained from NSE spectroscopy is modelled by the KWW function[40,41] namely:

$$\frac{S_{coh}(Q,t)}{S_{coh}(Q,0)} = A \exp\left[-\left(\frac{t}{\tau_{KWW}}\right)^{\beta_{KWW}}\right] \tag{6}$$

The fits shown in Fig. 4 lead to an amplitude $A$ less than unity. This indicates the presence of faster dynamics, which is not visible within the NSE time window, e.g., the EW or the fast $\beta$ process. For pure glycerol the fits were performed with the stretching parameter fixed to $\beta_{KWW} = 0.7$ s, following our previous studies[9] and the one by Wuttke $et\ al.$[43] However, a significant reduction of $\beta_\alpha$ is noticed upon addition of POSS: Here for 7.7 wt% POSS in glycerol, $\beta_\alpha$ varies between 0.63 and 0.49 for temperatures between 323 K and 273 K, respectively (Fig. 3). This increased stretching of the exponential decay under addition of POSS resembles the results from dielectric-spectroscopy recently reported by Johari and Khouri for different glass formers and POSS molecules where it was attributed to an increase of heterogeneity in the system[28,29]. Interestingly, the corresponding $\alpha$ relaxation time, $\langle \tau_\alpha \rangle$ only shows a slight increase (i.e., a slowing down of the $\alpha$ mode) when adding POSS as will be treated in more detail in the discussion section.

**Dielectric Spectroscopy (DS).** Figure 5 illustrates the frequency dependence of the dielectric loss $\varepsilon''(\nu)$ at four selected temperatures for 7.7 wt% of POSS in glycerol-$H_8$ (closed symbols). The data are compared with spectra for pure glycerol-$H_8$ (open symbols) taken from previous studies[3,4]. To enable a direct comparison of the relative EW amplitudes, the glycerol:POSS spectra were scaled to make the $\alpha$ peaks match those obtained for pure glycerol. Interestingly, this plot demonstrates that the addition of POSS to glycerol induces a sig-nificant increase of the amplitude of the EW, related to that of the $\alpha$ peak. A close inspection of the dielectric spectra published for diglycidyl ether of bisphenol-A (DGEBA), mixed with POSS molecules of about 2 nm size[28], reveals indications for a similar increase of the relative amplitude of the JG relaxation in this system.

POSS addition also seems to lead to a broadening of the $\alpha$ relaxation both at the low and high-frequency flanks of the peaks. However, at least at the high-frequency flank of the $\alpha$ peaks, this broadening may also arise from the stronger EW amplitude. To clarify this issue, fits of the spectra have to be performed. Figure 6 shows the dielectric constant (a) and loss spectra (b) at temperatures 195 K to 252 K for 7.7 wt% of POSS. The solid lines represent fits using the sum of a Havriliak-Negami (HN) function[48] (equation 7) for the $\alpha$-relaxation peak and a CC function[44] for the EW. The HN function is given by:

$$\varepsilon^* = \varepsilon_\infty + \frac{\varepsilon_s - \varepsilon_\infty}{[(1 + i\omega\tau)^{1-\alpha}]^\gamma} \tag{7}$$

Here, $\varepsilon_s$ is the static dielectric constant and $\varepsilon_\infty$ the high-frequency limit of the dielectric constant; the difference $\Delta\varepsilon = \varepsilon_s - \varepsilon_\infty$ represents the dielectric strength. Broadening and asymmetry are determined by the parameters $\alpha$ and $\gamma$ ($0 \leq \alpha < 1$; $0 < \gamma \leq 1$). For $\alpha > 0$ and $\gamma = 1$, equation (7) represents the symmetrically broadened CC function. For the low and high-frequency flanks of the relaxation peaks, equation (7) predicts power laws $\nu^m$ and $\nu^{-n}$, respectively, with $n = \gamma(1 - \alpha)$ and $m = 1-\alpha$. The fits were performed simultaneously for the dielectric constant $\varepsilon'(\nu)$ and $\varepsilon''(\nu)$. Here one should note that the justification for using a simple additive superposition of different contributions to the spectra may be doubted and alternatives were promoted[49]. However, usually the convolution approach proposed, e.g., in ref. 49 leads to similar results[50,51].

As revealed by Fig. 6, the model can successfully describe the complete spectra, in accord with the findings for pure glycerol and other glass formers[4,13,14]. The obtained EW-relaxation peaks, modelled by CC functions, are indicated by the dashed lines in Fig. 6(b).

There is also a corresponding contribution in the real part, indicated by dashed lines in Fig. 6(a). As shown in the inset of Fig. 6, it leads to a somewhat smoother rounding of the $\varepsilon'(\nu)$ curves when approaching $\varepsilon_\infty$ (ref. 8). In Fig. 3, the deduced exponents $n$ of the right flank of the $\alpha$ loss peaks are compared for pure glycerol[52] (open circles) and for the present glycerol-POSS mixture (closed circles). Obviously, within experimental resolution $n$ is identical for both materials. Thus, the fits do not reveal any significant variation of the high-frequency flanks of the $\alpha$ peaks by POSS addition and the apparent broadening in Fig. 5 can be completely ascribed to the stronger EW contribution.

One should be aware, however, that the data for pure glycerol, taken from ref. 52 (open circles in Fig. 3), were obtained from an analysis of the $\alpha$ peak alone, without taking into account the EW. Thus one may ask if the EW for pure glycerol also leads to an apparent broadening of the $\alpha$ peaks, similar to the findings for the glycerol/POSS mixture (Fig. 5). This would result in an artificial reduction of the values for $n$ reported in ref. 52, due to the superposition of the high-frequency flanks of the $\alpha$ peaks by the EW, instead of $n$ being a genuine property of the $\alpha$ relaxation. Therefore, in Fig. 3 we also include results for $n(T)$ obtained from fits[4] of the spectra of pure glycerol using the sum of a Cole-Davidson[53] (CD) and a CC function[44] (crosses), analogous to the present analysis of the glycerol:POSS data. They reasonably agree with those for the original evaluation without EW[52]. Obviously, in pure glycerol the amplitude of the EW is too small (in relation to the $\alpha$ relaxation) to lead to any significant apparent broadening of the $\alpha$ peak. Thus our statement that there is no evidence for any significant broadening at the right flank of the $\alpha$ relaxation of glycerol, caused by the addition of POSS, still holds.

In contrast to the high-frequency flanks, the performed fits reveal that the left flanks of the $\alpha$ peaks become shallower ($\alpha$ varies between 0.1 at 252 K and 0.21 at 195 K) under POSS addition if compared to pure glycerol where $\alpha$ is zero, i.e., $m = 1$ (refs. 3, 4, 52).

This also becomes obvious by the direct comparison of the spectra shown in Fig. 5. The fits in addition reveal a significant increase of the $\alpha$-relaxation times when 7.5 wt% POSS is added to glycerol (Fig. 7; see following section for a detailed discussion). The obtained relaxation strength varies between 26 (252 K) and 45 (195 K) and is smaller than in bulk glycerol ($\Delta\varepsilon = 45$ and 68 at the same temperatures, respectively[4]). A reduction of $\Delta\varepsilon$ compared to the pure compound was also found by Johari and Khouri in their studies of DGEBA, mixed with two different types of POSS molecules with a size of about 2 nm[28,29].

## Discussion

The analysis of the BS data presented in Fig. 2 illustrates the presence of two relaxation modes for $T \leq 294$K. Applying translational motion in a sphere and rotational motion along a sphere model to the EISF from the BS data reveals two different length scales, $R_{trans}$ and $R_{rot}$ (Fig. 2(d)), respectively. The length $R_{trans}$ decreases from 3 to 2 Å over $T$ = 413 - 294 K. It finally confines to 1.7 Å at 273 K, which is in close proximity to the hydrodynamic radius of the glycerol molecule ($R_H = 1.6$ Å)[36], obtained from viscosity data following the Stokes-Einstein relation. However, for $T \leq 294$ K a second length scale, $R_{rot}$, yields 1.16 – 1.0 Å. These values are close to, $R_{rot} = 0.8$ Å, and 0.96 Å, reported for pure glycerol[36] following DS and NMR data at $T > 1.3\ T_g$ ($\approx 190$ K). The small discrepancy is attributed to the incorporation of heterogeneity by the nanoparticles.

In refs. 28 and 29, the broadening of the $\alpha$ relaxation when adding POSS particles to glass-forming DGEBA was discussed in detail and stated to be compatible with the dynamic-heterogeneity interpretation of the commonly-found non-exponentiality of ultra-viscous-liquid dynamics. In Fig. 3, we compare the width parameters found for the present glycerol-POSS mixture using NS and dielectric experiments (closed symbols) with the results for pure glycerol (open symbols and crosses). As discussed in the preceding sec-tions, the NS data were analysed by the KWW function involving a width parameter $\beta_{KWW}$ (eq. (9)), and for the fits of the DS data of glycerol mixed with POSS the HN function had to be used (with width parameters $\alpha$ and $\gamma$; eq. (7)). Moreover, the DS data for pure glycerol[3,4] were analysed by the CD function[53] including a width parameter $\beta_{CD}$. The latter function leads to a qualitatively similar spectral shape as the Fourier-transformed KWW function, especially an asymmetric loss peak with exponents $m = 1$ at the left and $n = \beta_{CD} < 1$ at the right peak flank. To enable a comparison of the different data sets, Fig. 3 shows the exponent $n$ of the $\nu^n$ power law at the high-frequency side of the relaxation peak, which corresponds to $n = \beta_{KWW}$ for the KWW function, to $n = \gamma (1 - \alpha)$ for the HN function, and to $n = \beta_{CD}$ for the CD function. As discussed in detail in the preceding section, the analysis of the dielectric experiments (circles and crosses) does not evidence any significant variation of $n$ when POSS is added to glycerol. While the direct comparison of the spectra provided in Fig. 5 seems to suggest a broadening at the high-frequency flanks of the $\alpha$ peaks, the fits reveal that it is due to the superposition of the $\alpha$ peaks by the EW whose relative amplitude is markedly enhanced for the glycerol/POSS mixture. It should be noted that the invariance of $n$ under POSS addition, as deduced from dielectric spectroscopy in the present work, does not contradict the decrease of $\beta_{KWW} = n$ as reported in refs. 28 and 29 for the admixture of two different POSS compounds to DGEBA. In those works, only the $\alpha$-relaxation part of the spectra was analysed and a corresponding treatment of our data would reveal similar results. In any case, it should be noted that our data indeed indicate a clear broadening of the $\alpha$ relaxation when POSS is added, which, however, mainly seems to occur at frequencies $\nu < 1/(2\pi \langle \tau_\alpha \rangle)$ (cf. Fig. 5), which is not covered by a comparison of the parameter $n$.

Figure 3 also shows the results for the width parameters as obtained from the NS experiments. For NSE, data for both pure glycerol and glycerol with POSS are available (open and closed diamonds, respectively). $n$ for the pure compound partly deviates from the corresponding values from DS. However, as both methods sense different aspects of the glassy dynamics (DS: only reorientational, NS: primarily translational and partly reorientational) there is no principle reason for a match of both data sets. Indeed, in refs. 3 and 54, marked deviations in the $\alpha$-peak widths obtained by different experimental probes were reported for glycerol and two other glass formers. Interestingly, in contrast to the DS results, the width parameters determined from NSE for glycerol with POSS are markedly lower than for pure glycerol (Fig. 3, closed and open diamonds, respectively). This seems to imply that the translational dynamics is more sensitive to the heterogeneities induced by the POSS molecules than the reorientational one. In a naive picture one may rationalize this notion when assuming that the large POSS molecules partly are blocking the paths for the translational motions of the glycerol molecules, thus having a direct impact on this aspect of their dynamics. This obviously is not the case for the reorientational motions as detected by DS. Here a less direct influence of POSS on the heterogeneity sensed by the molecules can be assumed, which is caused by a variation of the interaction potentials with neighbouring molecules, no longer exclusively being glycerol molecules as in the pure compound. However, one should be aware that the determination of $n = \beta_{\mathrm{KWW}}$ from the NSE data is based on an evaluation with a single relaxation function without taking into account possible contributions from the EW. To account for the EW in the analysis, a larger density of $S(t)$ data points and higher precision would be necessary to reveal possible deviations from the KWW fits, expected at short times in Fig. 4. Similar considerations are valid for the BS data, also showing rather small values of $n$ in Fig. 3 (triangles), at first glance suggesting a POSS-induced increase of heterogeneity. As shown below, the BS data at low temperatures ($T \leq 294$ K) are clearly influenced or even dominated by the EW and it cannot be excluded that the EW also contributes to the measured data at higher temperatures.

We want to note that the considerations of the previous two paragraphs are based on the assumption that there are no significant contributions from reorientational motions associated to the POSS molecules. It is clear that additional loss peaks in the dielectric spectra, which would be expected for such additional processes, are not observed and the data can be well described by two relaxation functions only. However, small contributions from such processes, submerged under the dominating $\alpha$ relaxation cannot be excluded, which may hamper the precision of the determined width parameter of the $\alpha$ relaxation.

The deduced temperature-dependent average relaxations times $\langle \tau \rangle$ are shown in Fig. 7 in an Arrhenius plot. The solid circles represent the DS results for the $\alpha$-relaxation times, $\langle \tau_\alpha \rangle$, obtained in the present work for 7.7 wt% POSS concentration (for the HN function used in the fits, $\langle \tau_\alpha \rangle$ was approximated by the inverse loss-peak frequency, calculated[55] from $\tau$, $\alpha$, and $\gamma$). The published results for pure glycerol are shown as open circles[3,4]. A comparison reveals that the $\alpha$-relaxation time significantly increases by a factor of 2.5 - 7, depending on temperature, when POSS molecules are added to glycerol (the error of $\langle \tau_\alpha \rangle$ from DS is of the order of the symbol size). This agrees with the findings in ref. 29 for DGEBA mixed with 2 nm sized EcPOSS molecules (epoxycyclohexyl polyhedral oligomeric silsesquioxane). The increase of viscosity when admixing POSS molecules to glass-forming $m$-cresol, reported in ref. 34 is also in accord with a slowing down of the $\alpha$ relaxation. As noted in refs. 28 and 29, a simultaneous broadening and slowing down of the $\alpha$ relaxation, as found in the present work and for the DGEBA:EcPOSS system investigated in ref. 29, is in accord with the dynamic-heterogeneity scenario often assumed to govern glassy dynamics: Adding POSS molecules to a glass former of course should increase heterogeneity but also can be assumed to induce an increase of the typical length scale of heterogeneity, $\xi_{het}$. When further supposing a close relation of $\xi_{het}$ with $\xi_{CRR}$, the size of cooperatively rearranging regions, this should result in a slowing down of the $\alpha$ relaxation[28,56]. However, it should be noted that in ref. 28 a *de-* *crease* of $\alpha$-relaxation time was found when adding a different type of POSS (glycidyl polyhedral oligomeric silsesquioxane) to DGEBA than in ref. 29, which seems to contradict the above picture. A possible reason for this discrepancy may be the unclarified relation of $\xi_{het}$ and $\xi_{CRR}$.

The closed squares in Fig. 7 show the present DS results for the relaxation time of the secondary relaxation causing the EW, $\langle \tau_{EW} \rangle$. They are compared to two $\langle \tau_{EW} \rangle$ data sets for pure glycerol, taken from literature[4,14]. It is clear that the parameters of the excess wing, deduced from fits of the spectra with two relaxation functions, have high uncertainty due to the strong superposition of the EW relaxation by the dominating $\alpha$ relaxation. Within these uncertainties, no POSS-induced variation of $\langle \tau_{EW} \rangle$ is detected.

The solid and open diamonds in Fig. 7 represent the NSE data measured at $Q = 1.3$ Å$^{-1}$, which is close to the coherent static structure factor peak for pure glycerol ($Q_0 = 1.44$ Å$^{-1}$)[43]. A small increase of $\langle \tau_\alpha \rangle$ under POSS addition is revealed, just as for the dielectric data, but considering the shown error bars, this finding is of limited significance. The $\langle \tau_\alpha \rangle$ data from NSE seem to be systematically larger than those from DS. This may be a result of the different momentum transfer ($Q$), which is zero for DS. However, it should be noted that in literature $\langle \tau_\alpha \rangle$ is commonly found to *decrease* with $Q$ (see, e.g., refs. 57,58).

The open and solid triangles in Fig. 7 represent the $\langle \tau \rangle$ results of the glycerol-POSS mixture deduced from BS for $Q = 1.3$ and $1.5$ Å$^{-1}$ respectively, close to the coherent static structure factor peak. For the three highest investigated temperatures ($T \geq 323$ K; upright triangles), the $\langle \tau \rangle$ values from BS show reasonable agreement with those obtained from DS for pure glycerol. However, at lower temperatures (inverted triangles), the relaxation time obtained from the BS measurements deviates from $\langle \tau_\alpha \rangle$ from DS and at 274 K ($1000/T \approx 3.65$ K$^{-1}$) it agrees with $\langle \tau_{EW} \rangle$ determined by DS. Here the relaxation time is well within the instrument resolution[38] and it seems that the contribution from the EW starts to dominate within the BS time window. To our knowledge, this is the first time that the EW was detected by neutron BS investigations while there are several works where an EW or secondary relaxation were found by other NS methods[46,59–61]. On the other hand, the coherent NSE technique evidently is mainly sensitive to the $\alpha$-relaxation mode (cf. diamonds in Fig. 7). It should be noted that the obtained relaxation-time values at $\langle Q \rangle = 1.3$ Å$^{-1}$ and 1.5 Å$^{-1}$ are quite similar. Vispa et al. [ref. 61] have shown that at $Q = 1.3$ Å$^{-1}$ the data for incoherent NS matches with the DS data where the coherent contribution is negligible. The combination of our BS and NSE studies points to the fact that in close vicinity of the structural relaxation peak ($Q_0 = 1.44$ Å$^{-1}$) the coherent and incoherent dynamics follow similar trends although they are not quantitatively identical.

Another interesting result of the present work is the apparently much stronger relative amplitude of the EW in glycerol with POSS, compared to the pure compound (Fig. 5). This enhanced amplitude may also be the reason for its detectability in BS. It should be noted here that, based on the present results, it is not clear if indeed an increased relaxation strength $\Delta\varepsilon_{EW}$ causes this stronger EW or if its time scale becomes more separated from the $\alpha$ relaxation. In the fits of Fig. 6, the parameters $\Delta\varepsilon_{EW}$ and $\tau_{EW}$ obviously are strongly correlated because only the high-frequency flank of the EW relaxation is actually visible in the spectra. Interestingly, as reported by us earlier[56], putting ions into glycerol by admixing LiCl leads to a stronger EW, too. Moreover, the DGEBA-POSS mixtures investigated in ref. 28 also seem to show an increase of the relative amplitude of the secondary relaxation in this system, which is of JG type. Adding POSS or LiCl, undoubtedly leads to more heterogeneity in the glass-forming system. Thus, one may speculate about a heterogeneity-related mechanism affecting the EW. However, one should also be aware that the interactions between the molecules of the host glass former may be influenced by the added compound, leading to a variation of $\Delta\varepsilon_{EW}$ or of the $\tau_\alpha/\tau_{EW}$ ratio. For example, in ref. 56 the found effect of ion addition on the EW was discussed in terms of a partial breaking of hydrogen bonds, which may also occur when POSS is added to glycerol.

In summary, our investigations of glass-forming glycerol with 7.7 wt% POSS using DS and NS techniques have revealed strong effects of the added nanometre-sized heterogeneities on the structural dynamics and the excess wing of this molecular glass former. The most interesting findings are a simultaneous broadening and slowing down of the $\alpha$ relaxation. As pointed out by Johari[28,29], these effects are in accord with the heterogeneity scenario of glassy dynamics. Moreover, we find indications for an enhancement of the apparent relative amplitude of the EW. Performing similar experiments for other glass formers with EW or with well-pronounced JG relaxation seems a promising approach to learn more about the microscopic mechanisms leading to these only poorly understood features of glassy dynamics.

## Methods

The hybrid-plastic polyhedral oligomeric silsesquioxane (POSS) was bought from Hybrid chemical manufacturer (description: Al0136-Octa(3-hydroxy-3-methylbutyldimethylsiloxy) POSS with molecular weight = 1707.03). For SANS the solvent, deuterated (d) p-xylene ($C_6D_4(CD_3)_2$) was obtained from Sigma-Aldrich, with 99% isotopic purity for deuterium. For the incoherent neutron backscattering and dielectric spectroscopy, glycerol-$H_8$ ($C_3H_8O_3$) of $\geq 99.5\%$ purity from Sigma-Aldrich was used. For the coherent NSE experiments, deuterated glycerol-D8 ($C_3D_8O_3$) with a purity of 98% and 99% deuteration level was purchased from Cambridge Isotope Laboratories.

Accurately weighed amount of POSS and glycerol were added together inside a glove box in a glass vial, to avoid any moisture content. The glass vial was properly sealed with Teflon tape before taking it out of the glove box. The mixture was then homogenously mixed using a magnetic stirrer (with occasional vortex mixing) at an elevated temperature of 50°C for 1 week to make sure the solution is clear and homogeneous.

The small angle neutron scattering (SANS) studies were performed at the High Flux Isotope Reactor (HIFR), at ORNL using the Bio-SANS diffractometer. A typical SANS data reduction protocol, which consisted of subtracting scattering contributions from the empty cell (5 mm Hellma cells), background scattering, and sorting data collected from two different detector distances was used to yield normalized scattering intensities, $I(Q)$ (cm$^{-1}$) a.k.a. the macroscopic scattering cross-section ($d\Sigma/d\Omega$) as a function of the scattering vector, $Q$ (Å$^{-1}$). Data reduction was conducted employing the Igor Pro platform.

The quasi-elastic neutron scattering (QENS) experiments were performed at the Spallation Neutron Source (SNS), Oak Ridge National Laboratory (ORNL) using the BS spectrometer BASIS and the Spin Echo spectrometer SNS-NSE. BASIS is a time-of-flight (ToF) instrument where the energy exchanged by the neutron with the sample is measured by determining the time that each detected neutron takes to travel from the sample to the detector. The measurement configuration at BASIS was the standard one, with 3.5 $\mu$eV energy resolution (full-width at half maximum FWHM, $Q$-averaged value) and the dynamic range of ±120 $\mu$eV selected for the data analysis. The data presented in figure 2 were collected over a momentum transfer band, $\Delta Q$, 1.0 to 1.7 Å$^{-1}$, yielding an average $<Q> = 1.3\pm0.2$ and $1.5\pm0.2$ Å$^{-1}$. The sample was loaded into an annular cylindrical aluminum sample holder with an outer diameter of 29.0 mm and an inner diameter of 28.9 mm, resulting in a sample thickness of 0.05 mm. This sample thickness was chosen to keep the transmission of the incoming neutron beam ≥ 95%, a condition maintained for BS to avoid multiple scattering. The reduced BS data were normalized by an annular vanadium standard.

NSE spectrometers use the Larmor precession of the neutrons magnetic moment of an incoming polarized beam in the magnetic field of two large solenoids, one placed before and one placed after the sample, as an internal clock, thus measuring the energy exchanged be-tween the scattered neutron and the sample. This method reaches the highest energy resolution (neV) of all available neutron scattering instruments. At the SNS-NSE experiment an incoming wavelength band, $\Delta\lambda$, from 4.58 to 7Å was used with 42 time channels for the time-of-flight data acquisition. This allowed accessing a dynamic range of $5\ \mathrm{ps} \leq t \leq 38\ \mathrm{ns}$, at a fixed scattering angle ($2\theta = 71.59°$) over a momentum transfer band, $\Delta Q$, 1.05 to 1.59 Å$^{-1}$ of the $30 \times 30\ \mathrm{cm}^2$ detector. Due to the rather low scattering intensity of the sample, integration over all time channels was performed. This is equivalent to the use of an effective average incoming neutron wavelength of 5.76 Å at an average momentum transfer $<Q> = 1.3 \pm 0.2$ Å$^{-1}$. For the measured coherent NSE data, corrections were performed using resolution data from a TiZr sample and background from the empty cell. We used flat aluminum sample containers maintaining a sample thickness of 4 mm sealed with indium wires. The data reduction was performed with the standard ECHODET software package of the SNS-NSE instrument. For further details the reader is referred to ref. 47.

For the dielectric measurements, parallel plate capacitors (diameter 6 mm) were filled with the liquid sample material. The plate distance of 0.1 mm was fixed by corresponding glass-fibre spacers. The measurements were performed using a frequency-response analyser (Novocontrol Alpha-A analyser). Cooling was achieved by a closed-cycle refrigerator (CTI cryogenics).

α-relaxation and excess wing. *Chem. Phys.* **284,** 205–219 (2002).

## Acknowledgements

SG thanks R. Zorn for helpful discussions. We thank S. V. Pingali for helping us with the SANS experiment. Research conducted at ORNL's Spallation Neutron Source (SNS) and High Flux Isotope Reactor (HIFR) was sponsored by the Scientific User Facilities Division, Office of Basic Energy Sciences, US Department of Energy (DOE).

## Author Contributions

S.G., P.L. and M.O. designed the experimental studies. S.G., J.K.H.F and E.N. performed all the experiments and the analysis; S.G., N.J. and M.O. were responsible for planning and performing the neutron scattering measurements and data reduction. S.G, P.L, A.L. and M.O. wrote the paper; paper; all of the authors read the paper and provided editorial input.

## Additional Information

**Competing financial interests:** The authors declare no competing financial interests.

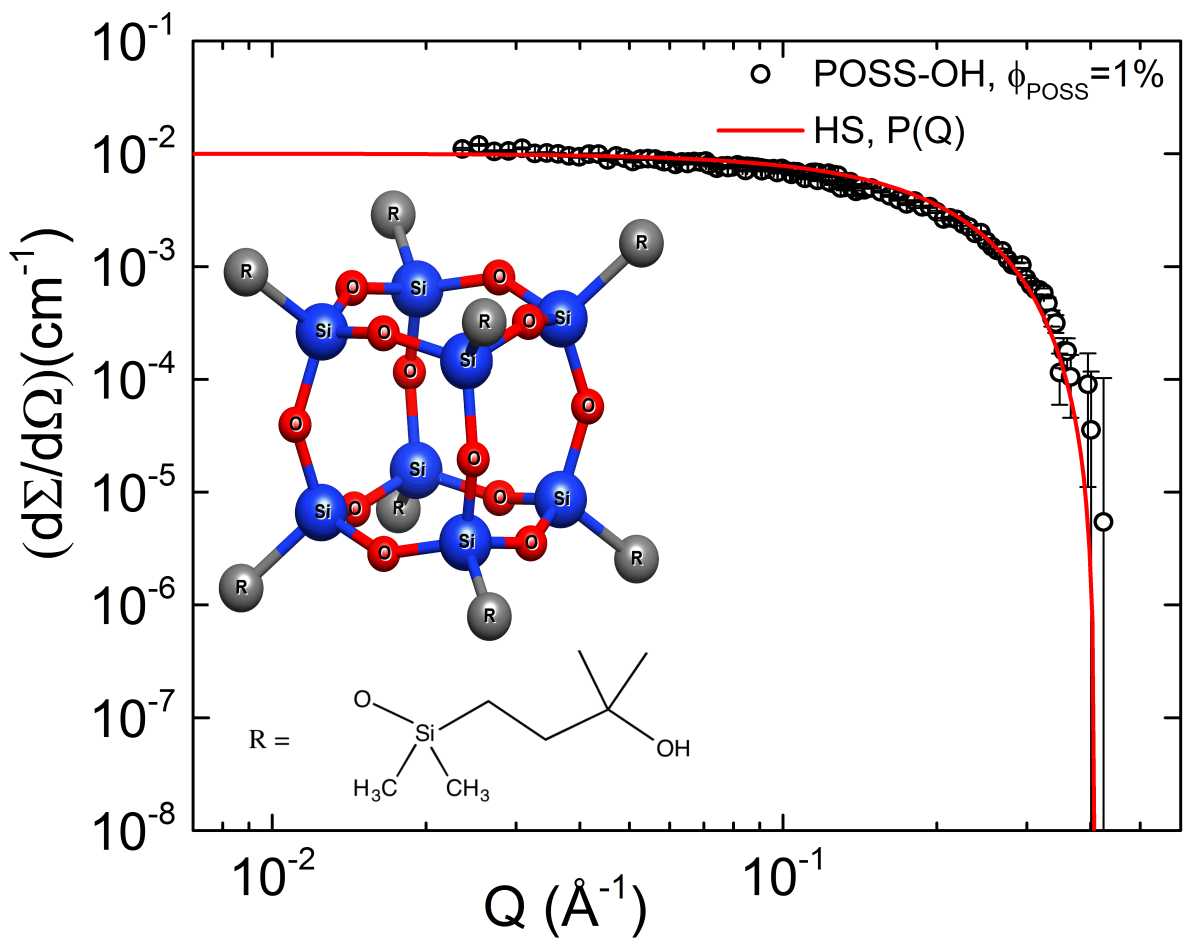

Figure 1. SANS scattering intensity of a deuterated p-xylene solution of POSS-OH with $\phi_{POSS} = 1\%$ volume fraction as a function of the scattering vector, $Q$. The solid line is a fit using a hard-sphere form factor (equation 1). Inset: Molecular structure of completely condensed POSS-OH[37]. The Si and O atoms in the silicon-oxygen cage are represented in blue and red, respectively.

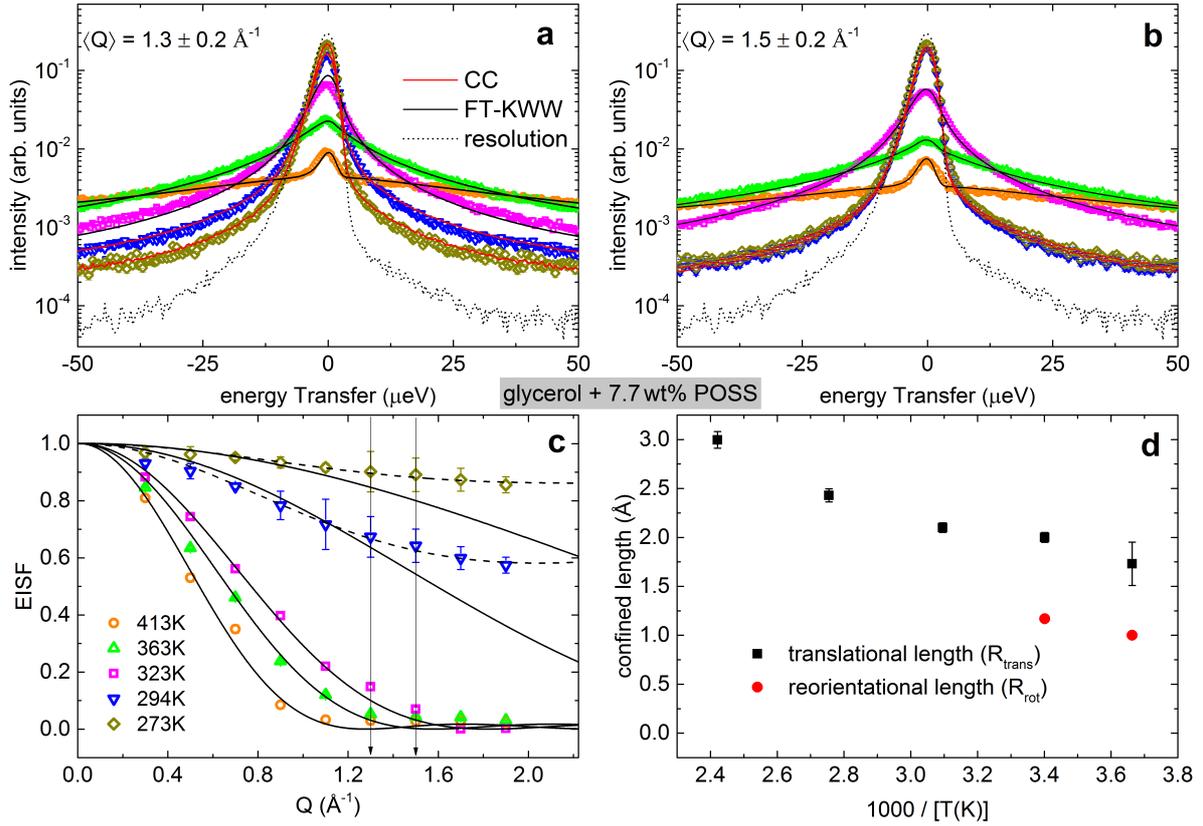

Figure 2. BS data in a semilogarithmic representation for 7.7 wt% POSS in glycerol as measured over a temperature range from 273 to 413 K at (a) $\langle Q \rangle = 1.3$ Å$^{-1}$ and (b) $\langle Q \rangle = 1.5$ Å$^{-1}$. The black solid lines are fits with equations (2) and (3), whereas the red solid lines are fits using equations (2) and (4) as explained in the text. The dashed lines indicate the resolution function, which is determined from the measurement of the sample at 20 K. (c) EISF from equation (2) as a function of Q. The solid lines represent the fits for diffusion in a sphere model and the dashed line represent the length scales in combination with the localized dynamics as explained in the text. (d) The confinement lengths for translational and reorientational dynamics as a function of inverse temperature.

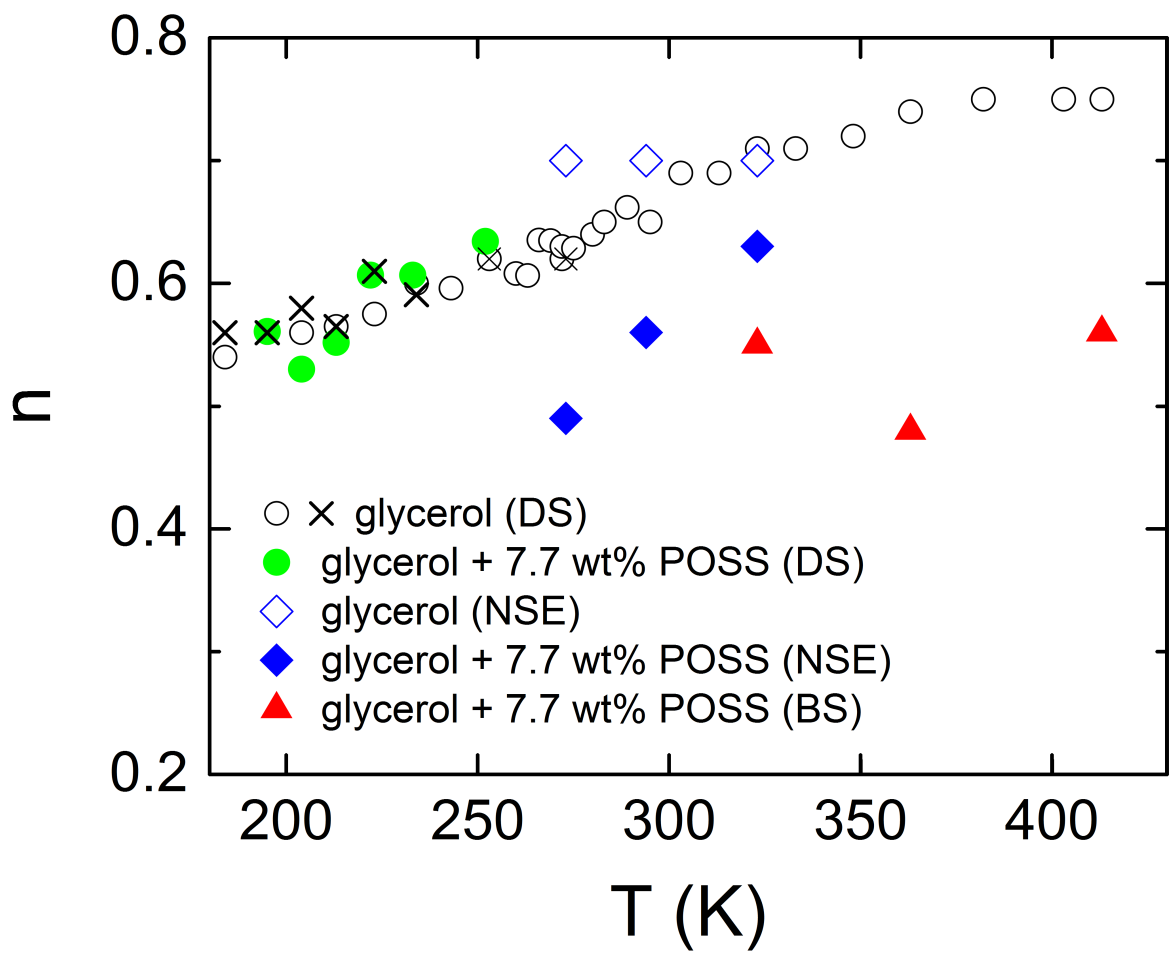

Figure 3. Width parameter *n* of the α relaxation for pure glycerol and glycerol with 7.7 wt% POSS as obtained from DS, NSE, and BS measurements. *n* is defined as the absolute value of the exponent, observed at the high-frequency flank of the α peak (see text). The DS data for pure glycerol were taken from refs. 3 and 52. The circles are from an evaluation of the α peak alone[52] while the crosses were obtained from fits with the sum of a CD and a CC function, also taking into account the EW[3].

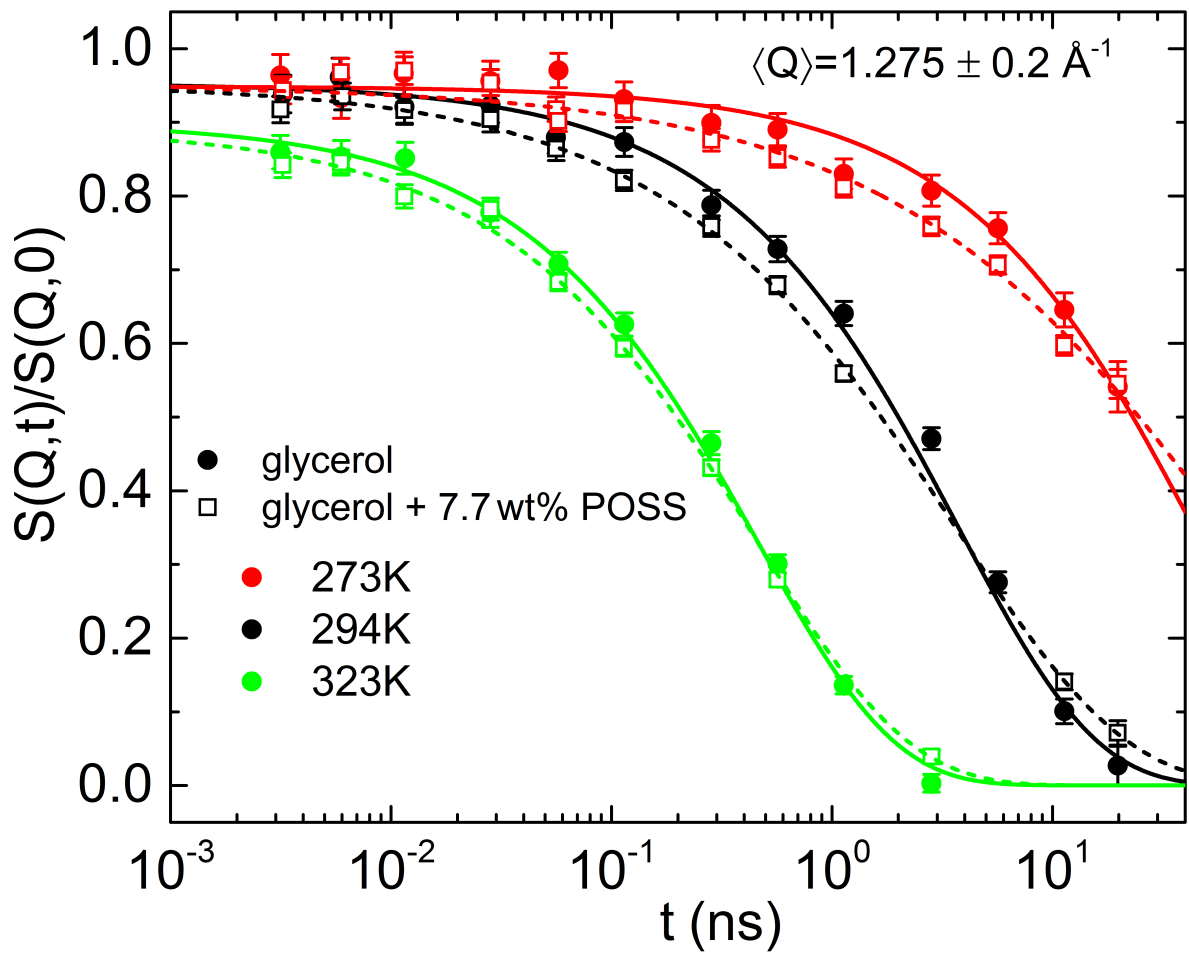

Figure 4. Normalized dynamic structure factor $S(Q,t)/S(Q,0)$ at $\langle Q \rangle = 1.3 \pm 0.2$ Å⁻¹ for pure glycerol (closed symbols) and glycerol with 7.7 wt% POSS (open symbols), measured at different temperatures as indicated in the legend. The lines are fits with the KWW function (equation 6).

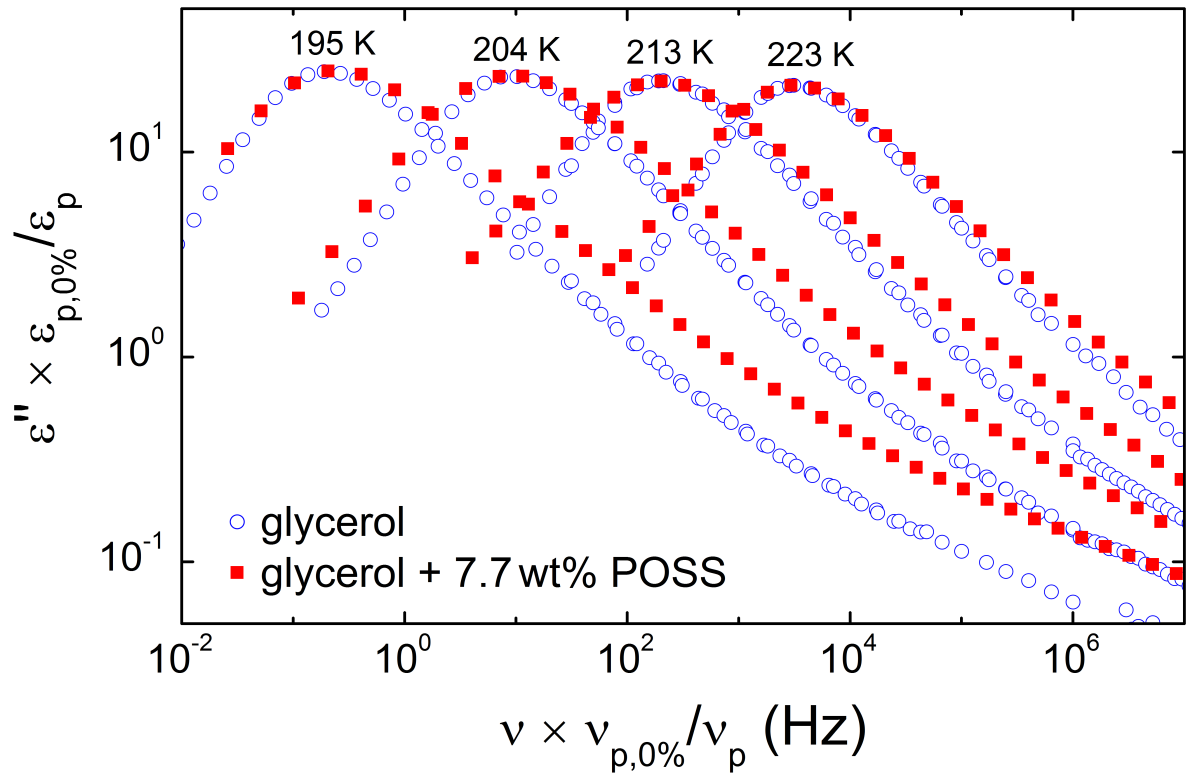

Figure 5. Dielectric loss spectra including the $\alpha$ relaxation peak and the EW region at $T$ = 195 K, 204 K, 213 K, and 223 K, measured for 7.7 wt% POSS in glycerol in comparison to the DS data for pure glycerol (from refs. 3,4). For the glycerol/POSS spectra, $\varepsilon''$ and $\nu$ were scaled to make the $\alpha$ peaks match those obtained for pure glycerol.

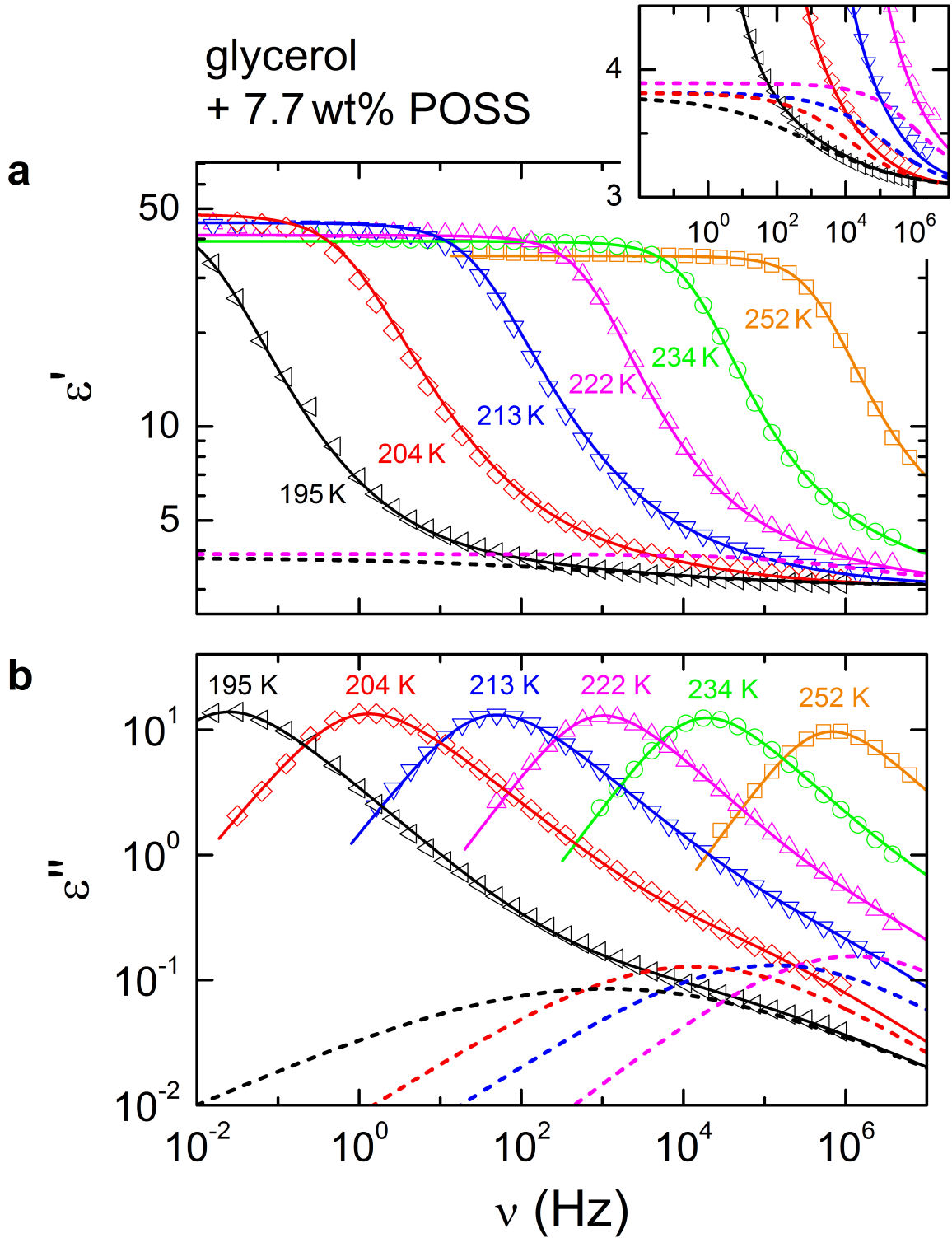

Figure 6. Spectra of dielectric constant (a) and loss (b) including the $\alpha$-relaxation peak and the EW region over a temperature range 195 – 263 K at 7.7 wt% POSS concentration in glycerol. The solid lines are fits with the sum of a HN function[48] (equation 7) for the $\alpha$ relaxation peak and a CC function[44] for the EW relaxation. The fits were performed simul-

taneously for $\varepsilon'(\nu)$ and $\varepsilon''(\nu)$. The dashed lines show the CC contributions for $T \leq 222$ K, leading to the EW as explained in the text (at higher temperatures, the EW is outside of the frequency window). For clarity reasons, in frame (a) the EW relaxation is shown for two temperatures only. The inset provides a zoomed view showing the EW contribution for all four temperatures.

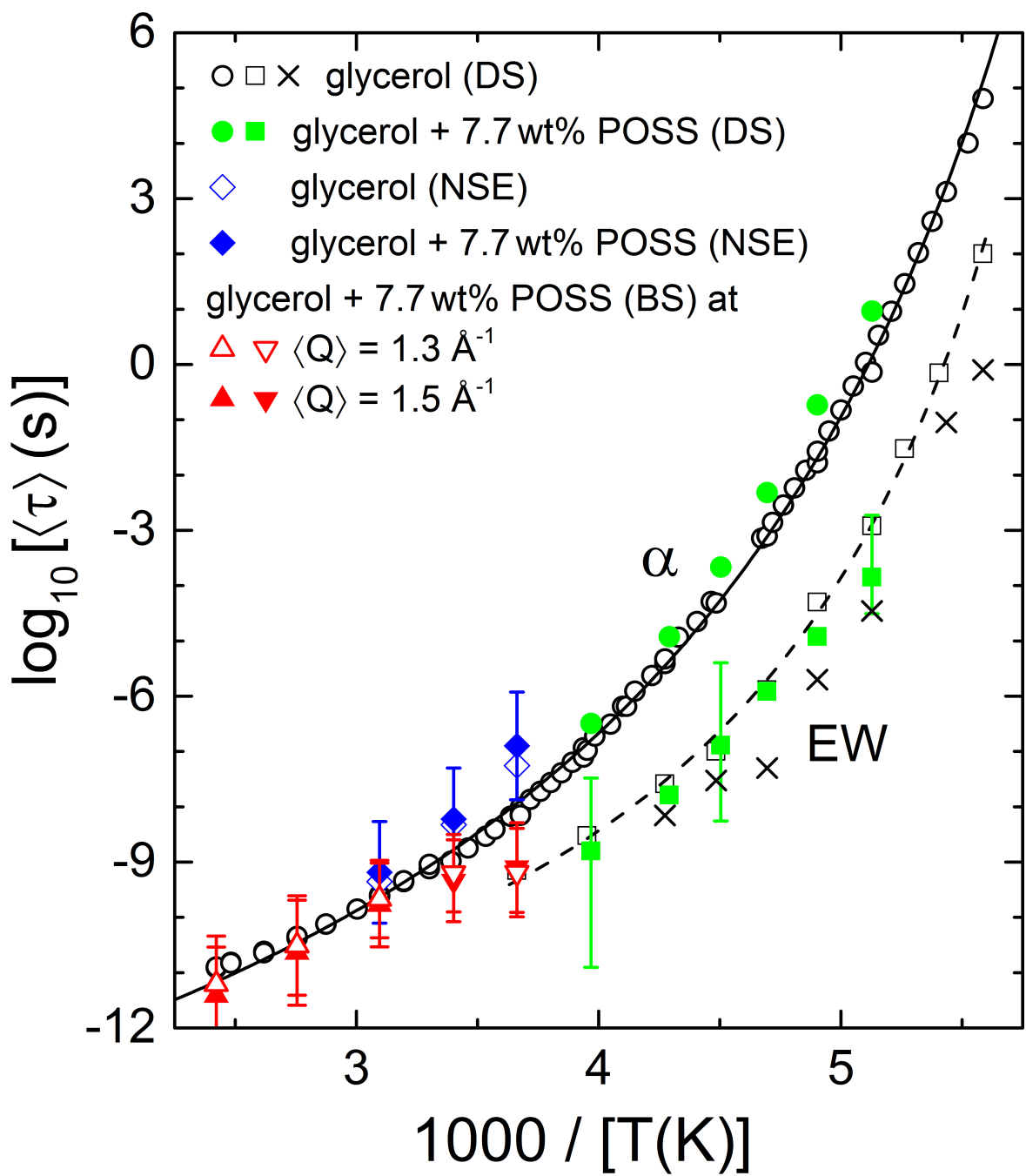

Figure 7. Arrhenius representation of the average relaxation times for glycerol mixed with 7.7 wt% POSS in comparison to pure glycerol[3,4,14]. Open circles show the average $\alpha$-relaxation times of pure glycerol while the open squares and crosses represent the EW-relaxation times, both determined by DS[3,4,14]. The open squares were obtained from an evaluation using the sum of a KWW and a CC function[4] while the crosses were deduced assuming the sum of a CD and a CC function[14]. The $\alpha$-relaxation times represented by the open circles[3,4] were determined from fits of the $\alpha$ peaks alone, leading to practically identical results as the fits combining two peak functions to account for the EW. The closed circles and squares denote the $\alpha$-relaxation and EW-relaxation times, respectively, as determined in the present work for 7.7 wt% POSS in glycerol from fits of the DS spectra with the sum of a HN and CC function (Figure 6). The upright and inverted triangles represent the relaxation times determined from the BS data (Fig. 2) at $\langle Q \rangle = 1.3$ and 1.5 Å$^{-1}$, for 7.7 wt% POSS in glycerol. The open and solid diamonds represent the relaxation times obtained from the NSE data (Fig. 4) at $\langle Q \rangle = 1.3 \pm 0.2$ Å$^{-1}$ for pure glycerol and for 7.7 wt % POSS in glycerol, respectively. The solid and dashed lines are fits of the open circles and squares with the Vogel-Fulcher-Tammann function[62,63].